\documentclass[runningheads]{llncs}
\usepackage{makecell}
\usepackage{graphicx}
\usepackage{amssymb}
\usepackage{framed,multirow,color}
\usepackage{marvosym}
\begin{document}

\title{Degradation-invariant Enhancement of Fundus Images via Pyramid Constraint Network}

%
\titlerunning{PCE-Net: Pyramid Constraint Network for Fundus Image Enhancement}
%
%
%
\author{
Haofeng Liu\inst{1,2} \and 
Heng Li\inst{2,}\textsuperscript{\Letter} \and 
Huazhu Fu\inst{3} \and 
Ruoxiu Xiao\inst{4} \and 
Yunshu Gao\inst{1,2} \and 
\\Yan Hu\inst{2,}\textsuperscript{\Letter} \and 
Jiang Liu\inst{1,2,5} 
}
\authorrunning{H. Liu, H. Li, H. Fu, et al.}
\institute{
Guangdong Provincial Key Laboratory of Brain-inspired Intelligent Computation, Southern University of Science and Technology, Shenzhen, China\and
Department of Computer Science and Engineering, Southern University of Science and Technology, Shenzhen, China\\
\email{lih3, huy3@sustech.edu.cn}\and
IHPC, A*STAR, Singapore\and
The School of Computer and Communication Engineering, University of Science and Technology Beijing, China
\and Singapore Eye Research Institute, Singapore National Eye Centre, Singapore
}
\maketitle              

\begin{abstract}
As an economical and efficient fundus imaging modality, retinal fundus images have been widely adopted in clinical fundus examination.
Unfortunately, fundus images often suffer from quality degradation caused by imaging interferences, leading to misdiagnosis.
Despite impressive enhancement performances that state-of-the-art methods have achieved, challenges remain in clinical scenarios.
For boosting the clinical deployment of fundus image enhancement, this paper proposes the pyramid constraint to develop a degradation-invariant enhancement network (PCE-Net), which mitigates the demand for clinical data and stably enhances unknown data. 
Firstly, high-quality images are randomly degraded to form sequences of low-quality ones sharing the same content (SeqLCs).
Then individual low-quality images are decomposed to Laplacian pyramid features (LPF) as the multi-level input for the enhancement.
Subsequently, a feature pyramid constraint (FPC) for the sequence is introduced to enforce the PCE-Net to learn a degradation-invariant model.
Extensive experiments have been conducted under the evaluation metrics of enhancement and segmentation.
The effectiveness of the PCE-Net was demonstrated in comparison with state-of-the-art methods and the ablation study.
The source code of this study is publicly available at https://github.com/HeverLaw/PCENet-Image-Enhancement.
\end{abstract}

\section{Introduction}
Due to the convenience, economy, and safety, fundus photography has widely served as a routine clinical examination. Fundus photography is taken by a specialized low-power microscope with an attached CCD (Charge Coupled Device) camera designed to photograph the interior of the eye, including the retina, optic disc, macula, and posterior pole (i.e., the fundus)~\cite{zhang2022machine,ZHANG2022104037}. 
Unfortunately, collecting fundus images under non-ideal light conditions is a frustrating experience. 
The causation of image quality degradation includes imperfections in the fundus camera optics, aberrations of the human eye, improper camera adjustment, flash lighting, or focusing during the exam. 
Therefore fundus observation is prone to be impacted by the degradation, resulting in the uncertainty of diagnosis. 

Efforts have been long made to enhance the quality of fundus images~\cite{li2021applications,liu2022domain}. 
Contrast limited adaptive histogram equalization (CLAHE) and Fourier transforms have been collaborated to restore degraded fundus images~\cite{mitra2018enhancement}. 
Inheriting the guided image filtering (GIF), Cheng et al.~\cite{cheng2018structure} designed a structure-preserving guided retinal image filtering (SGRIF) to correct the fundus images from cataract patients. 
With filtering technique and root domain in frequency, Cao et al.~\cite{cao2020retinal} proposed an enhancement method for the retinal images.
More recently, with the advance of deep learning techniques, fundus image enhancement has achieved state-of-the-art performances. 
Segmentation networks were introduced to preserve retinal vessels in enhancement~\cite{shen2020modeling}. 
Based on unpaired image translation~\cite{zhu2017unpaired,park2020contrastive}, importance-guided semi-supervised contrastive constraining (I-SECRET)~\cite{cheng2021secret} as well as structure and illumination constrained GAN (StillGAN)~\cite{ma2021structure} were proposed to learn the enhancing mappings from unpaired data. 
Considering the gap between synthesized and clinical data, domain adaptation~\cite{li2021Restoration,li2022annotation} was leveraged to generalize the restoration model from synthesized data to real ones by accessing the clinical data. 

Despite existing enhancement methods performed decently in the laboratory, challenges remain in the clinical scenarios.
i) Due to the complexity of clinical imaging interference, it is impractical for the model to traverse every possible degradation in training.
ii) Unpaired data tends to drive enhancement models to abandon some fundus characteristics (such as vessels and abnormality), while synthesized data probably result in performance-drop on real data. 
iii) As the foundation of fundus assessment, retinal structures need to be more efficiently preserved in the enhancement.
iv) Access to clinical data is still necessary for domain adaptation to bridge the gap between synthesized and clinical domains.

To circumvent the above issues, a degradation-invariant fundus enhancement method, termed PCE-Net, is developed in this paper via pyramid constraint. 
SeqLCs are generated from identical images as the training data with content consistency.
Then, the LPF is employed to preserve the retinal structures in the enhancement, and FPC is proposed to constrain the consistency of embeddings for enforcing the learning of a degradation-invariant model. 
The main contributions of this paper are summarised as follows:
\begin{itemize}
    \item For boosting the enhancement of clinical fundus images, the PCE-Net is developed to learn the degradation-invariant enhancement model from simulated data.
    \item The SeqLC and FPC are married to enforce the training of a degradation-invariant model, and the LPF is employed to preserve retinal structures.
    \item Three evaluations are presented in the experiment to verify the effectiveness of the PCE-Net, and superior performance to state-of-the-art algorithms is achieved in the enhancement of clinical fundus images.
\end{itemize}

\section{Methodology}
\begin{figure}[tbp]
\begin{centering}
\includegraphics[width=0.98\textwidth]{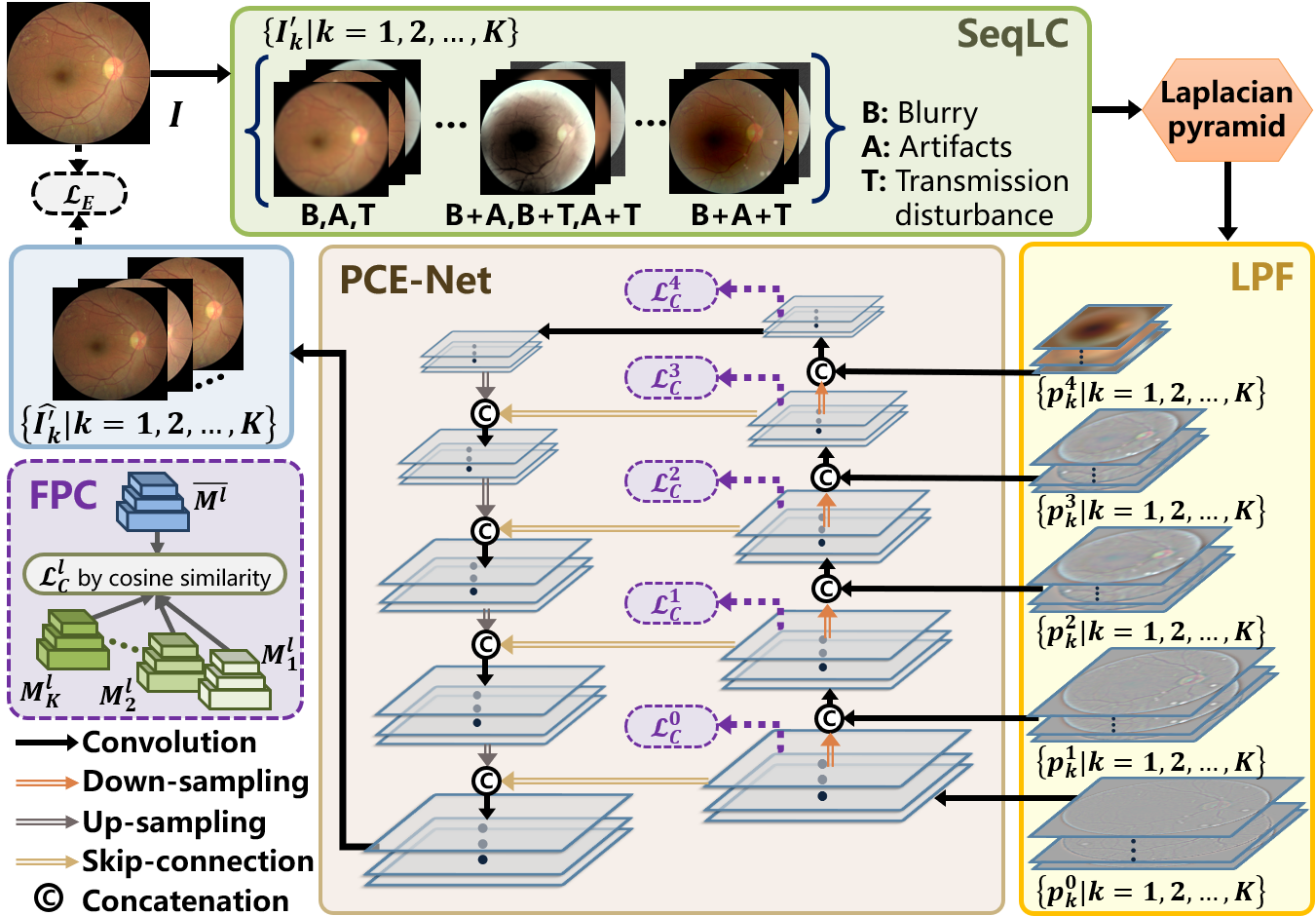}
\par\end{centering}
\caption{Overview of PCE-Net. Given a high-quality image $I$, the SeqLC $\{I'_k|k=1,2,..., K\}$ is randomly synthesized. Then $I'_k$ is decomposed as the LPF $\{p_k^l|l=0,1,...,L\}$ to boost structure-preserving in PCE-Net. Subsequently, the FPC is quantified by the consistency loss $\mathcal{L}_{C}^l$ at each encoding layer to enforce the model to learn degradation-invariant representations.} \label{fig:overview}
\end{figure}

As exhibited in Fig. \ref{fig:overview}, SeqLC, LPF, and FPC are implemented in the proposed PCE-Net. 
Specifically, the SeqLC is randomly degraded from an identical high-quality image for the training of the enhancement network. 
The LPF is adopted to forward multi-level inputs to boost structure-preserving of the enhancement network. 
The FPC leads the network to learn the degradation-invariant enhancement model.

\subsection{Sequence of Low-quality Images with the Same Content}
\label{paragraph:RD}
In clinics, fundus images often suffer from quality degradations caused by imaging interferences~\cite{shen2020modeling}. 
Unfortunately, the enormous cost of collecting high-low quality fundus image pairs and the limited volume of available data prevent the development of enhancement models.
To collect training data and boost the learning of degradation-invariant models, high-quality images are randomly degraded to generate low-quality ones with the same content.
SeqLCs are thus acquired to import content consistency into the model training.

As demonstrated in the green block of Fig. \ref{fig:overview}, the SeqLC $\{I'_k|k=1,2,..., K\}$ is generated from an identical high-quality fundus image $I$ to enforce the learning of the degradation-invariant model using the content consistency. 
The three imaging interferences modeled in~\cite{shen2020modeling}, including blurry, artifacts, and transmission disturbance, are randomly combined to degrade $I$.
Accordingly, $\{I'_k|k=1,2,..., K\}$ is formed by the low-quality images with various degradations but the same content.

Furthermore, denoting the enhancement results of $\{I'_k|k=1,2,...,K\}$ as $\{\widehat{I'_k}|k=1,2,...,K\}$, the enhancement loss $\mathcal{L}_{E}$ under the supervision from $I$ is defined as:
\begin{equation}
\mathcal{L}_{E}=\mathbb{E}\left[\frac{1}{K} \sum_{k=1}^{K}\left \| I-\widehat{I'_k} \right \|_{1}\right].
\label{eq:lr}
\end{equation}

\subsection{Laplacian Pyramid Features}
Laplacian Pyramid decomposes an image into a linear invertible image representation of multi-level frequency features and has been extensively used to boost the structure information in image generation~\cite{denton2015deep}, super-resolution translation~\cite{liang2021high}, and semantic segmentation~\cite{ghiasi2016laplacian}.
Considering retinal structures are essential in clinical fundus examination, Laplacian Pyramid is introduced to constrain retinal structures in the enhancement model.

Laplacian Pyramid is implemented with down-sampling and up-sampling operations. 
For an image $I$ with the size of $s \times s$, the down-sampling operation $\delta(\cdot)$ is carried out by smoothing with a Gaussian filter and then down-sampling to $\frac{s}{2}\times\frac{s}{2}$.
Subsequently, the up-sampling operation $\mu(\cdot)$ is performed by resizing the image back to $s \times s$.
To acquire an $L$-level Laplacian pyramid feature (LPF) from an image $I$, a Gaussian pyramid $\mathcal{G}(I)=\{g^l|l=0,1,...,L\}$ is firstly computed, where $g^0=I$ and $g^n$ denotes the output of $I$ underwent $n$ times $\delta(\cdot)$. 
Then the LPF $\mathcal{P}(I)=\{p^l|l=0,1,...,L\}$ is achieved, where $p^l=g^l-\mu(g^{l+1})$ and $p^L=g^L$.

As shown in the yellow block of Fig. \ref{fig:overview}, an LPF with the $L$ of 4 is extracted for retinal structure preservation in the proposed PCE-Net. 
LPFs $\{\mathcal{P}(I'_k)|k=1,2,...,K\}$ are calculated from $\{I'_k|k=1,2,..., K\}$, and then the LPFs are forwarded to the network by concatenating with the output of the corresponding layers. Denote $f^l$ as the output from the $l$-th layer in PCE-Net, the output of the next layer is given by
\begin{equation}
f_k^{l+1}=\mathrm{Conv}([p_k^{l+1}, f^{l}_k]), l=0,1...,L-1, k=1,2,...,K,
\label{eq:fl}
\end{equation}
where $[\cdot]$ is the concatenation operation, $\mathrm{Conv(\cdot)}$ refers to the convolution operation at each layer, and $f^0_k=\mathrm{Conv}(p_k^0)$.

\subsection{Feature Pyramid Constraint}
The underlying content consistency in SeqLC enables the learning of an enhancement model invariant to degradations. 
Concretely, for a degradation-invariant enhancement model, corresponding consistency should share in the features embedded from $\{I'_k|k=1,2,...,K\}$.
However, straightly constraining the numerical consistency of the feature maps in PCE-Net is too inflexible to converge at the optimal solution.
Therefore, the FPC is proposed to boost the enhancement model training effectively.

The FPC replaces the numerical consistency constraint to learn the representations invariant to degradations, which leads to a degradation-invariant model.
Concretely, spatial pyramid pooling (SPP)~\cite{2015he}, a multi-scale pooling operation, is employed to quantify the FPC. 
Denote SPP as $\sigma(\cdot)$, where the pooling scales of $2\times 2$, $4\times 4$, and $8\times 8$ are conducted.
The feature map extracted by SPP is defined as $M^l_k=\sigma(f^l_k)$, where $f^l_k$ denotes the output at the $l$-th layer in PCE-Net from $I'_k$.
Subsequently, as exhibited by the purple blocks in Fig.~\ref{fig:overview}, the feature maps extracted with SPP are adopted to calculate the consistency loss $\mathcal{L}_{C}^l$ at each encoding layer of PCE-Net using cosine similarity:
\begin{equation}
\mathcal{L}_{C}^l=\mathbb{E}\left[ \frac{1}{K}\sum_{k=1}^{K} \left(1-\frac{M^l_k\cdot \overline{M^l}}{\|M^l_k\| \|\overline{M^l}\| }\right)\right],
\label{eq:lcl}
\end{equation}
where $\overline{M^l}=\frac{1}{K}\sum_{k=1}^{K}{M^l_k}$. 
The FPC is hence given by
$\mathcal{L}_{C}=\sum_{l=0}^{L} \mathcal{L}_{C}^l$.

Accordingly, the total loss function is defined as:
\begin{equation}
\mathcal{L}_{total}=\mathcal{L}_{E}+\lambda_C \mathcal{L}_{C},
\label{eq:loss}
\end{equation}
where $\lambda_C$ is the weight to balance the total loss and set to 0.1 in training.

\section{Experiments}
\subsection{Implementation Details}
To demonstrate the effectiveness of the proposed PCE-Net, a comparison and ablation study were implemented in three evaluations, including enhancement evaluations with and without reference, and vessel segmentation.

\begin{table}[htbp]
\scriptsize
\centering {\caption{Datasets used in the experiments}
\label{tab:dataset} }%
\renewcommand{\arraystretch}{1.3}
\setlength\tabcolsep{4pt}
\begin{tabular}{|p{1.7cm} |p{1.8cm} |p{4cm}  |p{1.8cm}| }
\hline
Evaluation & Training set & Test set & Metrics\\
\hline
Full reference enhancement & \multirow{3}{1.8cm}[-8pt]{Good subset in EyeQ: 16,817 high-quality fundus images} & FIQ: 196 low-high quality fundus image pairs &  SSIM, PSNR\\
\cline{1-1} \cline{3-4}
Non-reference enhancement & & Usable \& Reject subset in EyeQ test dataset: 6,435 mediocre- and 5,540 low-quality fundus images & FIQA, WFQA\\
\cline{1-1} \cline{3-4}
Segmentation & &  FIQ & IoU, DSC\\
\hline
\end{tabular}
\end{table}

As reported in Table~\ref{tab:dataset}, the public dataset of EyeQ~\footnote[1]{https://github.com/HzFu/EyeQ} and a private fundus dataset FIQ were used in the experiments. 
Six state-of-the-art enhancement methods, including Cao et al.~\cite{cao2020retinal}, pix2pix~\cite{isola2017image}, CycleGAN~\cite{zhu2017unpaired}, I-SECRET~\cite{cheng2021secret}, CofeNet~\cite{shen2020modeling}, Li et al.~\cite{li2021Restoration} were compared with PCE-Net.
The 'Good' subset in EyeQ was employed as the high-quality images to construct paired data for the model training. 
High-low quality image pairs are contained FIQ, serving as the test data for the full reference evaluation of enhancement and segmentation.
The non-reference evaluation was tested on the 'Usable' and 'Reject' subsets in EyeQ.
The enhancement performance was quantified by structural similarity (SSIM) and peak signal-to-noise ratio (PSNR)~\cite{hore2010image} in the full reference evaluation, while by fundus image quality assessment (FIQA)~\cite{cheng2021secret} and weighted FIQA (WFQA) in the non-reference one.
FIQA was defined as the ratio of the test images predicted as 'Good' by MCF-Net~\cite{fu2019evaluation}, and WFQA was calculated by assigning the weights of \{2, 1, 0\} to the images predicted as \{'Good', 'Usable', 'Reject'\}.
The segmentation was measured by the metrics of intersection over union (IoU) and Dice score (DSC)~\cite{jadon2020survey}. The segmentation results were obtained from enhanced images and the reference using a U-Net trained on the DRIVE dataset.

In the experiments, a U-Net was designed to construct PCE-Net illustrated in Fig.~\ref{fig:overview}.
Note that low-quality images are individually degraded from high-quality ones without the SeqLC.
The model trained 200 epochs with Adam optimizer, which the learning rate was 0.001 for the first 150 epochs and decayed gradually to 0 for the following 50 epochs. 
The batch size was 16, and instance normalization was applied. The input image was resized to $256\times256$.
The same training strategy and augmentation data were applied to all algorithms.

\subsection{Comparison and Ablation Study}
Comparison of enhancement and segmentation with the state-of-the-art algorithms is visualized in Fig.~\ref{fig:comparison}. And quantitative evaluations of the comparison and ablation study are summarized in Table~\ref{tab:comparison}.

\subsubsection{Comparison}
\begin{figure}[tbp]
\begin{centering}
\includegraphics[width=11.9cm]{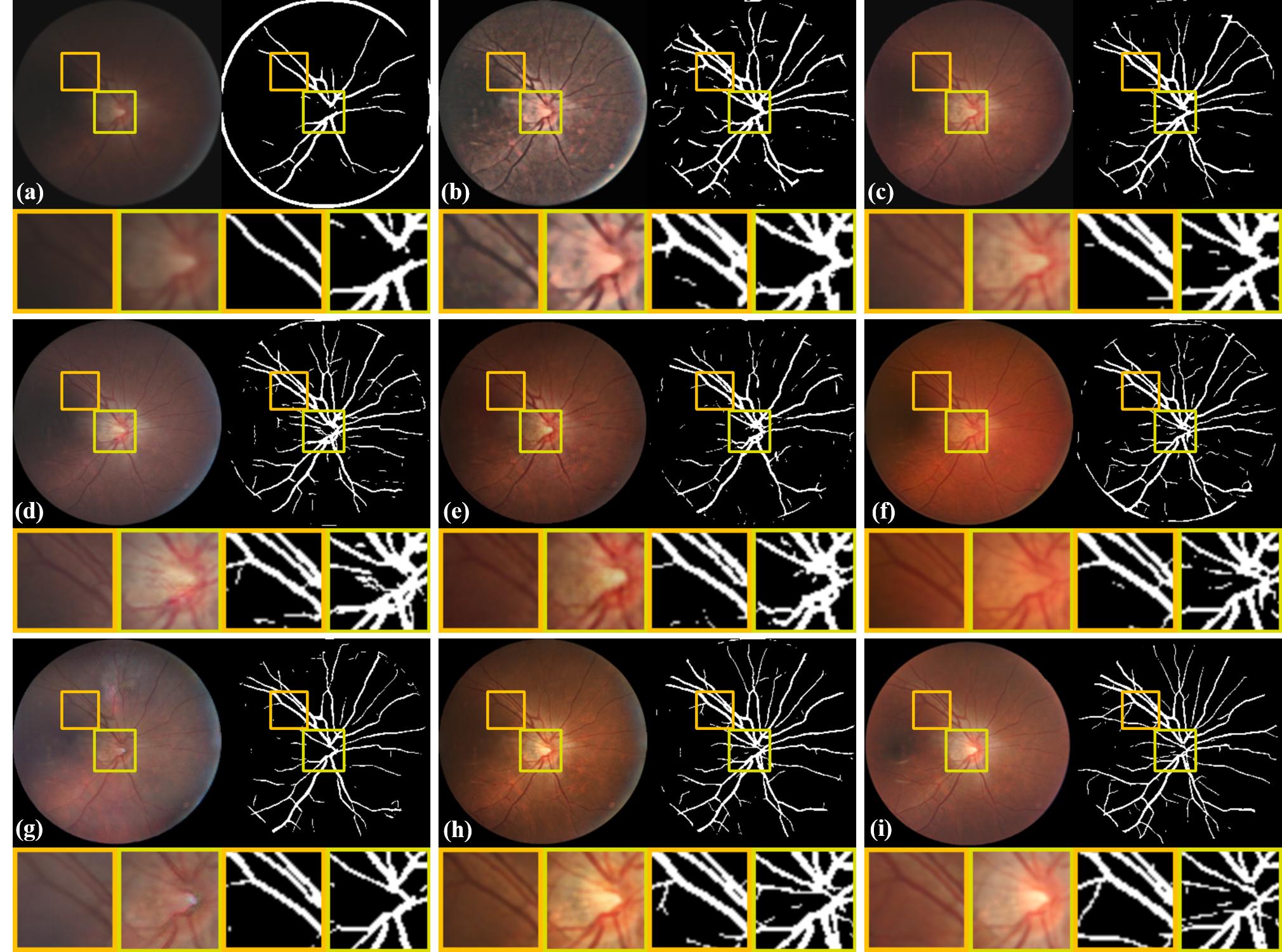}
\par\end{centering}
\caption{
Visual comparison on the fundus image with full reference. PCE-Net outstandingly enhances the clinical degraded fundus image and outperforms the state-of-the-art methods. (a) Low-quality real image. (b) Cao et al.~\cite{cao2020retinal}. (c) pix2pix~\cite{isola2017image}. (d) CycleGAN~\cite{zhu2017unpaired}. (e) I-SECRET~\cite{cheng2021secret}. (f) CofeNet~\cite{shen2020modeling}. (g) Li et al.~\cite{li2021Restoration}. (h) PCE-Net (ours). (i) Reference.} 
\label{fig:comparison}
\end{figure}

As illuminated in Fig.~\ref{fig:comparison} (a), as a result of the quality degradation, it is difficult to observe the fundus, and the segmentation is also severely impacted.
Based on image filtering, Cao et al.~\cite{cao2020retinal} improved the image clarity in Fig.~\ref{fig:comparison} (b), but the enhancement result was apparently different from the common ones.
Due to the grievous interferences, the detailed structures are abandoned in the enhancement by pix2pix~\cite{isola2017image}.
Thanks to training on unpaired images, CycleGAN~\cite{zhu2017unpaired} and I-SECRET~\cite{cheng2021secret} are more convenient to deploy in clinical scenarios.
However, their preservation of retinal structures is limited by the unpaired data, such that artifacts appear in the enhancement and segmentation result of Fig.~\ref{fig:comparison} (d) and (e).
Using vessel annotations as the structure guidance, promises CofeNet~\cite{shen2020modeling} remarkable preservation for fundus vessels, but leads to neglect of the unannotated structures.
To suppress the gap between the synthesized and real data, Li et al.~\cite{li2021Restoration} introduces domain adaptation to generalize the enhancement model from synthesized images to real ones. 
Despite the advance in the performance of Fig.~\ref{fig:comparison} (g), access to real data is required in the model training.
Though only learning from synthesized data, superior performance on clinical image enhancement is presented by PCE-Net in Fig.~\ref{fig:comparison} (h).
An enhanced image with prominent structure visualization is provided by PCE-Net, and the corresponding segmentation result is highly consistent with that of the high-quality reference in Fig.~\ref{fig:comparison} (i).

Table~\ref{tab:comparison} summarizes the quantitative performances in the comparison.
As the image clarity is improved by Cao et al.~\cite{cao2020retinal}, decent segmentation is achieved. However, the appearance different from common images leads to inferior performance in the enhancement evaluation.
Although the image-to-image translation methods of pix2pix~\cite{isola2017image} and CycleGAN~\cite{zhu2017unpaired} present reasonable performances in the enhancement evaluation with reference, the mediocre results in the non-reference evaluation and segmentation demonstrate their limitations on enhancing retinal structures.
Because an importance map and contrastive constraint are introduced in I-SECRET~\cite{cheng2021secret} to boost structure preservation and local consistency, advantages in the enhancement and segmentation are exhibited.
CofeNet~\cite{shen2020modeling} and Li et al.~\cite{li2021Restoration} respectively employ vessel segmentation and edge detection as structure guidance, achieving reasonable segmentation performance.
Nevertheless, as a restoration method of cataract fundus images, the performance of Li et al.~\cite{li2021Restoration} has been impacted in the enhancement of degraded images.

Owing to the degradation-invariant representation and multi-level structure preservation, PCE-Net achieves outstanding quantification results in each evaluation compared to the state-of-the-art methods.

\begin{table}[tbp]
\scriptsize
\centering
\caption{
Comparisons and ablation study of PCE-Net on evaluations of full-reference, non-reference enhancement, and segmentation. Clear and degraded denote the high-quality images and the low-quality ones before enhancement. N.A. denotes the results were not available. PCE-Net achieves superior performance in the enhancement and segmentation.
}
\label{tab:comparison} 
\renewcommand{\arraystretch}{1.2}
\begin{tabular}{p{4cm} | p{1.0cm}<{\centering} p{1.0cm}<{\centering} | p{1.0cm}<{\centering} p{1.0cm}<{\centering}| p{1.0cm}<{\centering} p{1.0cm}<{\centering}}
\hline
& \multicolumn{2}{c|}{Full reference}  &\multicolumn{2}{c|}{Non-reference}&\multicolumn{2}{c}{Segmentation}\\
\cline{2-7}
 &SSIM & PSNR  & FIQA & WFQA & IoU & DSC \\ 
\hline
Clear & N.A. & N.A. & 0.95 & 1.95 & N.A. & N.A.\\
Degraded & 0.766 & 16.55 & 0.09 & 0.67 & 0.350 & 0.518 \\
\hline
\hline
Cao et al.~\cite{cao2020retinal}& 0.741 & 20.17 & 0.27 & 0.68 & 0.431 & 0.602\\
pix2pix~\cite{isola2017image}& 0.846 & 21.96 & 0.18 & 0.87 & 0.420 & 0.592\\
CycleGAN~\cite{zhu2017unpaired}& 0.844 & 21.36 & 0.25 & 0.90 & 0.420 & 0.591 \\
I-SECRET~\cite{cheng2021secret}& 0.841 & 21.85 & 0.43 & 1.18 & 0.423 & 0.593 \\
CofeNet~\cite{shen2020modeling} & 0.857 & 22.41 & 0.45 & 1.26 & 0.427 & 0.599\\
Li et al.~\cite{li2021Restoration} & 0.839 & 21.88 & 0.37 & 1.15 & 0.435 & 0.606\\
\hline
\hline
PCE-Net w/o SeqLC, LPF, FPC & 0.849 & 21.46 & 0.11 & 0.73 & 0.400 & 0.572\\
PCE-Net w/o LPF, FPC  & 0.854 & 22.35 & 0.24 & 0.99 & 0.424 & 0.596\\
PCE-Net w/o FPC  & 0.860 & 22.55 & 0.50 & 1.32 & 0.447 & 0.618\\
PCE-Net (ours) & \textbf{0.871} & \textbf{23.09} & \textbf{0.55} & \textbf{1.38}  & \textbf{0.464} & \textbf{0.634}\\
\hline
\end{tabular}
\end{table}

\subsubsection{Ablation study}
The ablation study is also provided in Table~\ref{tab:comparison} to verify the effectiveness of the proposed modules. 
The SeqLC not only prepares sufficient training data for the enhancement model, but also imports content consistency for enforcing the degradation invariance.
By decomposing the images into multi-level inputs to the network, the LPF boosts the structure preservation in the enhancement and thus performs remarkably in the enhancement and segmentation.
Based on the underlying consistency from the SeqLC, the model in training is constrained by the FPC to learn degradation-invariant representations.
Consequently, a prominent enhancement model invariant to various degradations is acquired with the proposed modules.

\section{Conclusion}
To address the impact of quality degradation on clinical fundus examination, a fundus image enhancement network via pyramid constraint, called PCE-Net, is proposed to construct the degradation-invariant model from generated data.
The SeqLCs are first formed by low-quality images degraded from identical high-quality ones for training the enhancement model. Then the degraded images are decomposed into the LPFs to boost the retinal structure preservation of the enhancement model. 
Furthermore, FPC is proposed to enforce the learning of degradation-invariant representations based on the underlying content consistency in the SeqLCs.
The qualitative and quantitative experiments in enhancement and vessel segmentation demonstrate that PCE-Net presents superior performance compared to state-of-the-art methods. And the ablation study verifies the effectiveness of the proposed modules.
\section*{Acknowledgment}
This work was supported in part by Basic and Applied Fundamental Research Foundation of Guangdong Province (2020A1515110286), The National Natural Science Foundation of China (8210072776), Guangdong Provincial Department of Education (2020ZDZX3043), Guangdong Provincial Key Laboratory (2020B121201001), Shenzhen Natural Science Fund (JCYJ20200109140820699, 20200925174052004), and A*STAR AME Programmatic Fund (A20H4b0141).

\bibliographystyle{splncs04}
\bibliography{mybibliography}
\section*{Appendix}
\begin{figure}[htbp]
\begin{centering}
\includegraphics[width=0.93\textwidth]{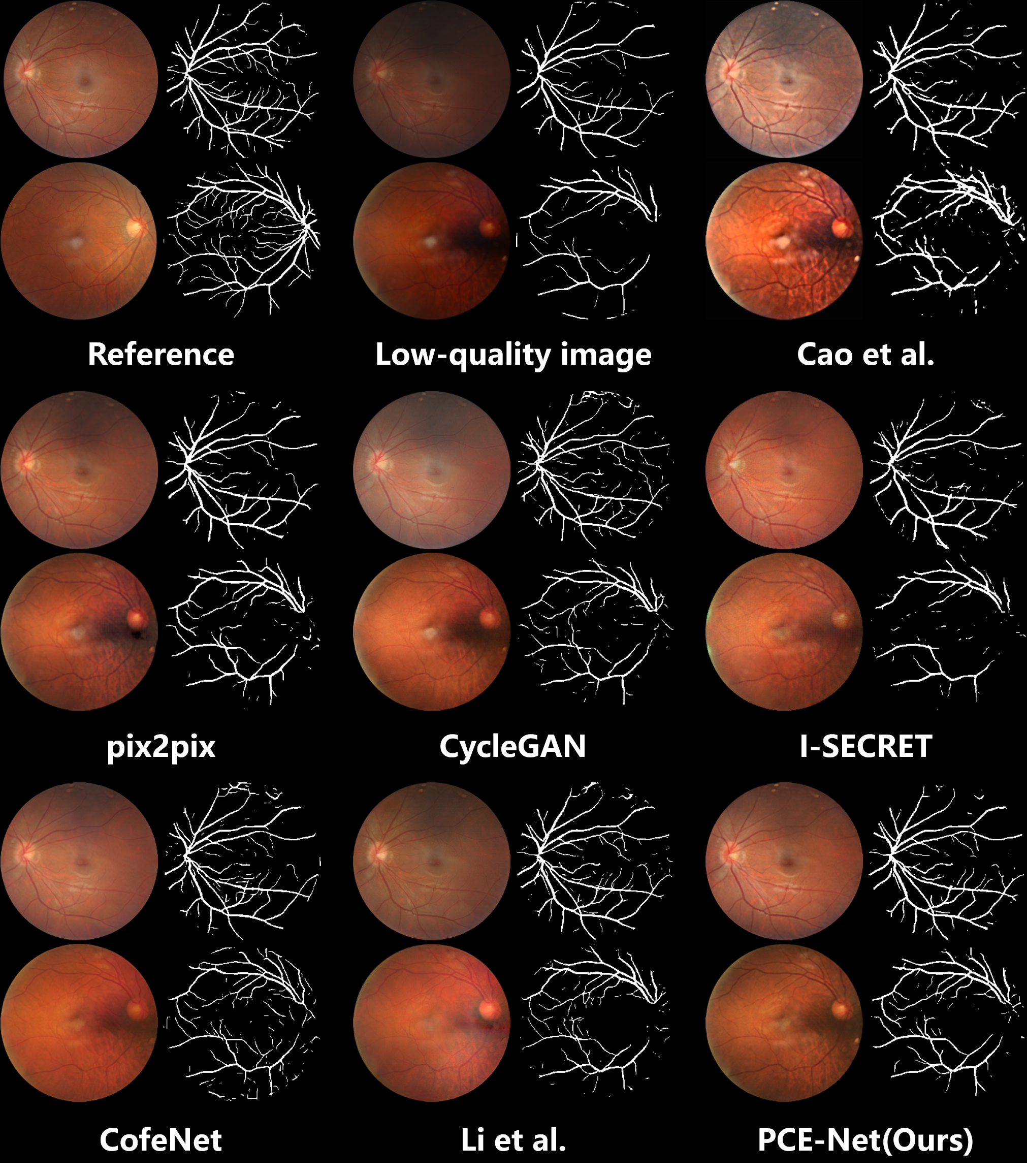}

\par\end{centering}
\caption{
Visual comparison on the fundus images from FIQ with full reference. PCE-Net outstandingly enhances the clinical low-quality fundus images, demonstrating its enhancement advantage.} 

\label{fig:appendix1}
\end{figure}

\begin{figure}[htbp]
\begin{centering}
\includegraphics[width=0.93\textwidth]{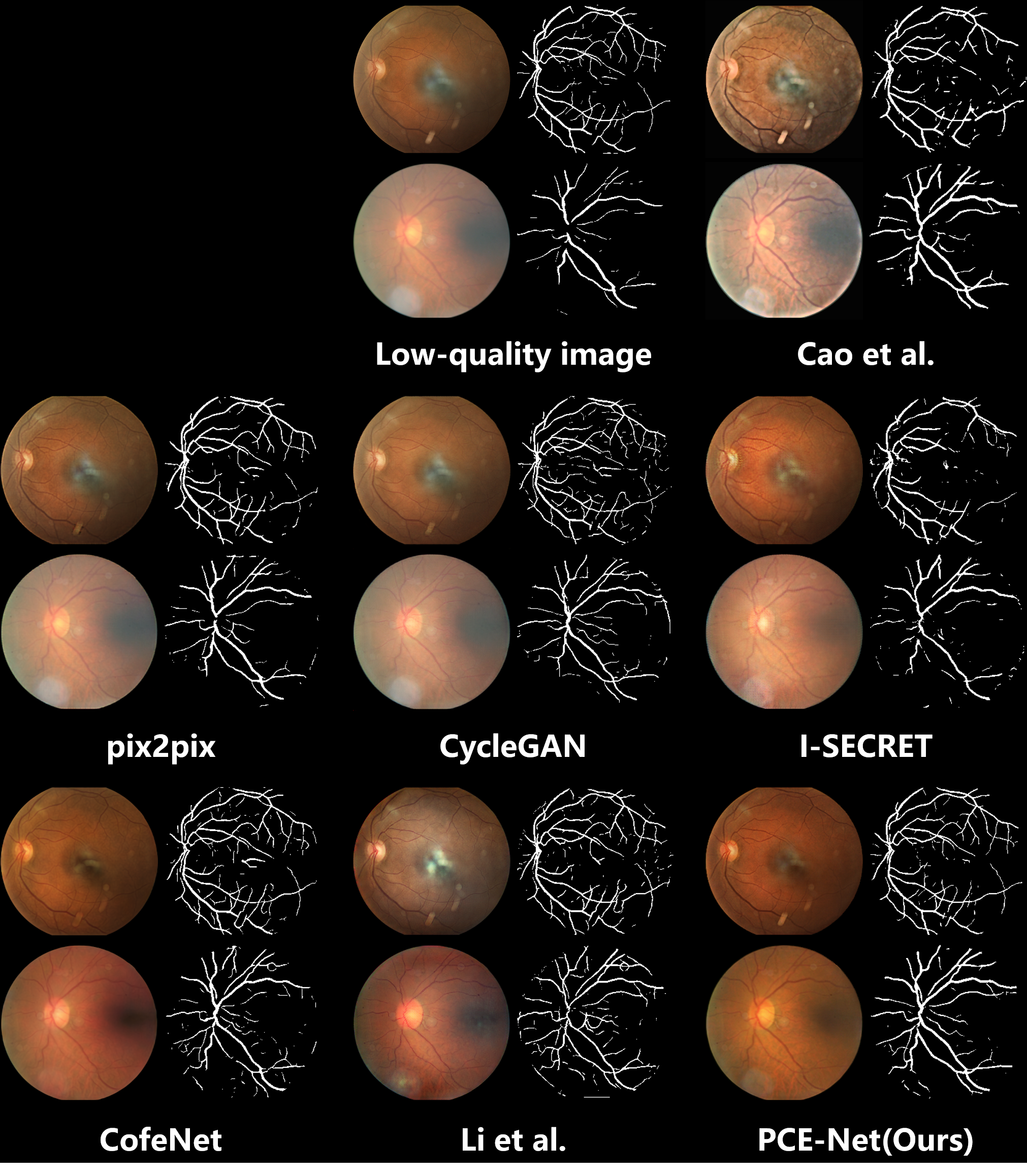}
\par\end{centering}
\caption{
Visual comparison on the fundus images from EyeQ without reference. PCE-Net provides remarkable visualization results, leading to superior vessel segmentation.} 
\label{fig:appendix2}
\end{figure}

Two additional enhancement examples are respectively provided for  FIQ and EyeQ. The enhancement and segmentation results of FIQ are provided in  Fig. \ref{fig:appendix1} and those of EyeQ are in Fig. \ref{fig:appendix2} is from EyeQ.

\end{document}